# Geotechnical problems related with loess deposits in Northern France

P. Delage[1], Y.J. Cui[1] & P. Antoine[2]

[1]Ecole Nationale des Ponts et Chaussées, CERMES, Paris, France
[2]Laboratoire de Géographie Physique, UMR CNRS 8591, Meudon, France

**Abstract**

Special problems were encountered in some areas in Northern France where the high speed railways (TGV Nord) crossed some loess deposits that appeared to be specially sensitive to change in water content and susceptible to collapse. Numerous sinkholes appeared along some sections of the line following wet climatic periods. After a general geological and geotechnical presentation of loess deposits and collapse susceptibility, in which some tools of the mechanics of unsaturated soils are reconsidered with special application to loess collapsibility, this paper describes the results of a geotechnical study carried out on block samples of intact loess. Collapse susceptibility is examined in the light of microstructure observation. The dependence of collapse to water content changes is examined and the validity of various existing collapse criteria is investigated.

*Keywords:* loess, collapse, unsaturated, criterion, microstructure.

## 1 Introduction

The Northern area of France is located in a key position between Paris and Northern European countries, including Belgium and the United Kingdom. With increasing European economic exchanges, various infrastructures have been constructed in this area, including highways, high speed train railways (TGV), gas-pipe lines,… The TGV high speed railway has an important role in the transportation of individuals to UK, Belgium, Germany, Netherlands and other countries.

Northern France is characterised by the presence of large loess deposits that are relatively well known on a geological point of view. However, some specific geotechnical problems have been faced in some areas where some loess deposits appeared to be quite sensitive to water content changes and susceptible to collapse. Similar collapsible loess are well known in Danubian countries, China, North America and Russia (Abelev &Abelev (1979), among others countries. Although Northern France has a rather wet climate, the geological disposition of the loess deposits and the strcuture of loess profiles is such that some layers of loess remain significantly unsaturated, even during the wet Autumn and Winter periods.

Collapsible unsaturated loess deposits have a low plasticity index and a high porosity. The sensitivity to changes in water content appeared to be particularly critical in the case of high speed train railways, where tolerances in terms of allowable displacements are more severe than in standard railways or in other transportation infrastructures.

More specifically, numerous sinkholes appeared during the railways construction and the beginning of the TGV service in 1993, in zones where the layer of vegetal soil had been removed and where the layer of collapsible loess was directly exposed to climatic changes and water precipitation. Among 43 sinkholes observed during that period all along the 50 km long section where the railway went through the collapsible loess deposits, it appeared after careful examination that 19 were of a natural origin. The remaining 21 were considered as artificial, in relation with holes and trenches excavated by soldiers in that area where fights were particularly intense during World War I (Somme battle among others). More recently, many other sinkholes have been detected following the two rainy periods that occurred during Winter 2001 and Spring 2002.

In parallel with the implementation of in-situ stabilisation methods, various investigations have been developed to better understand the origin and possible evolution of these sinkholes. Also, the sensitivity of the loess to train vibrations (20 Hz frequency) is presently under investigation.

This paper deals with a detailed geological characterisation of loess deposits and with a detailed geotechnical study carried out on a typical loess profile. Special attention is also paid to susceptibility to collapse and to the identification of relevant criteria in order to help anticipating similar problems in new in-situ survey related to other projects in similar deposits.

## 2 Geological description of the loess of Northern France

Loess is a silty aeolian sediment, transported in periglacial conditions and deposited in cold steppe environments, near the margins of main Quaternary ice-sheets, mainly around 50°N in the Northern hemisphere. Some deposits also exist in South America.

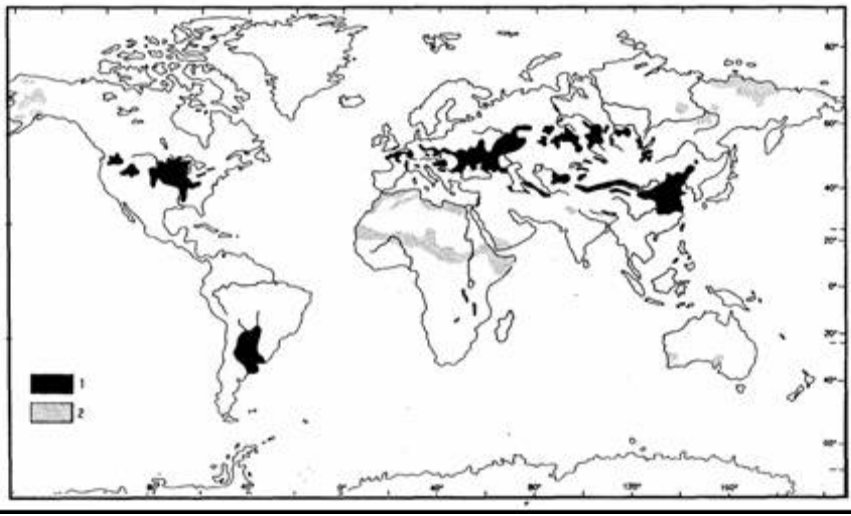

Figure 1. Loess deposits in the World (after Pecsi 1990)

P. Delage, Y.J. Cui, P. Antoine

Figure 1 presents the geographical repartition of loess deposits all over the world. In the Figure, number 1 (black areas) represents loess and number 2 (grey areas) represents "loess like" sediments. Loess deposits are located on plateaux, slopes and in main alluvial basins. In China, the thickness of the layer of "Plateau loess" can locally reach 300 m. Loess deposits are also found on the plateau of Siberia, in the Russian lowlands and in the basins of the rivers Danube, Rhine and of the Mississippi river in North America and in the Argentina Pampa.

Figure 2 presents the mechanisms of formation of aeolian loess deposits :

- thin particles produced by glacial abrasion are washed out, transported by proglacial flows and redeposited close to the moraine (outwash plains)
- sand, silt and clay particles that are submitted to freeze-thaw cycles are eroded and transported by violent and permanent cold and dry winds generated by high pressures above ice-sheets
- larger sand particles are first deposited as duns and as sand cover
- silt and clay particles are transported towards low pressure zones and high atmosphere before being deposited due to changes in wind regime : fall in wind speed, obstacle, capture by herbaceous vegetation or snow cover.

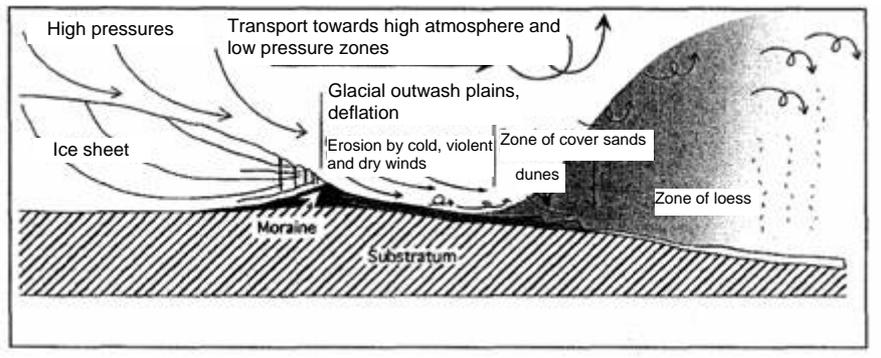

Figure 2. Scheme of the deposition of aeolian loess in the proximity of large ice-sheets (after Andersen & Borns 1997)

Figure 3 shows a typical cross section of the valley of the Somme river in Northern France, showing local mechanisms of loess deposition during the beginning of the Upper Weischel period (20-25000 BP).

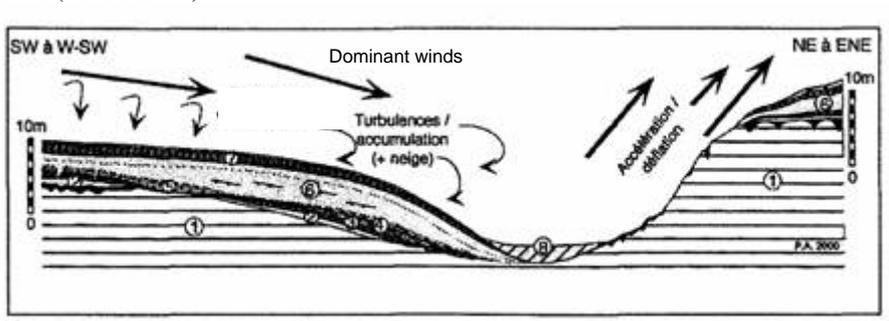

Figure 3. Local deposition mechanisms through a typical cross section of the Somme valley in Northern France (Antoine 2002)

In this area at that period, dominant winds were coming from South West to West South West (see also Figure 4). The substratum in this area is typically composed of chalk with upper clayey weathering. The aeolian loess deposit (number 6 in the figure) typically occurs on the west side of the valley that was protected from wind due to turbulences. Loess deposit was also enhanced by snow. On the other hand, the opposite side of the Valley was submitted to wind erosion.

In Northern France, loess is widespread from northern Brittany to Normandy and North, except on steep slopes exposed to W or NW and on the narrow ridges of the Bocage normand (Sommé 1980, Lautridou 1985, Lebret and Lautridou 1991, Antoine et al. 2003). These regions are characterised by a continuous or sub-continuous loess cover, which is a very important component of the present-day landform (about 3 to 5 m thick in Picardy and Northern France, and locally up to 8 m in Northeast France). The typical loess is also up to 8m thick on the gentle leeside slopes (protected from the dominant NW-NNW winds) of asymmetrical valleys, especially in Picardy. The main loess accumulations occur in the Lower Seine valley, the eastern Somme basin and in the North. Over the whole area the main typical calcareous units are attributed to the Weichsel Upper Pleniglacial, between 15 and 25 000 years BP.

Figure 4 shows the extension of the Last Glacial loess in Northern France, together with the direction of dominant winds. The TGV railway is also schematically represented, together with the area where sensitive loess deposits were found.

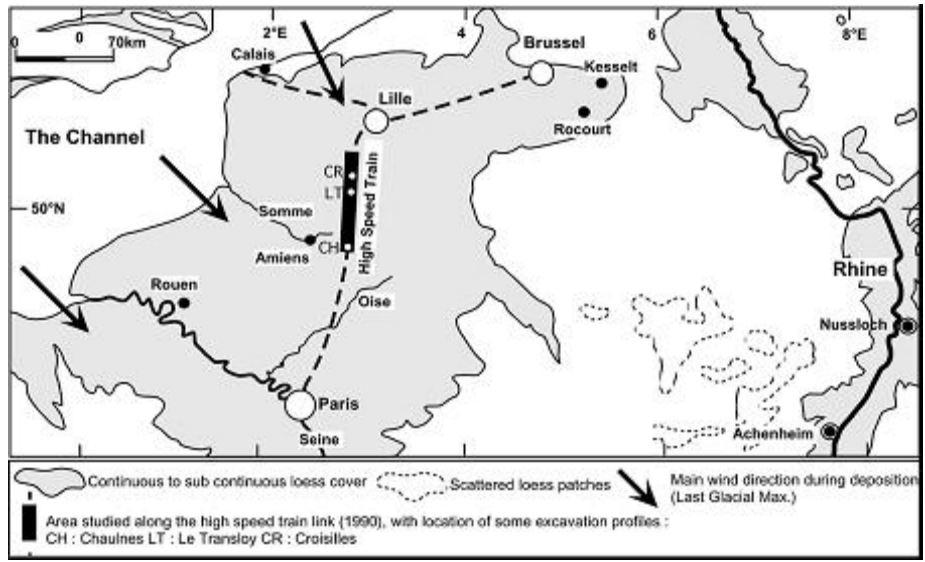

Figure 4. Extension of the Last Glacial loess in Northern France (Antoine 2002, modified)

In the regions characterised by a discontinuous and thin cover, the loess as accumulated particularly on leeside slopes, overlying fluvial gravels in valleys such as that of the Orne river (near Caen) and the Somme river, and at the junction between slopes and alluvial or marine terraces (see the mechanism presented in Figure 3).

In Northern France, the main loess facies is a typical homogeneous yellowish-grey, porous calcareous loess ($CaCO_3$ up to 18%), locally including chalk granules. Other facies, occurring on the plateaux of Upper Normandy and Picardy is a non-calcareous "limon à doublets", characterised by a sharp alternation of millimetric brown clay rich layers and yellowish-grey clay poor layers (post depositional feature).

According to grain size gradients, to the morphology of the main loess units, and to mineralogy (presence of heavy minerals), the source of the loess was located in the English Channel during periods of low sea level (Lautridou, 1985, see Figure 4). Aeolian reworking of calcareous detrital sediments exposed on the Channel floor produced large amounts of silt, which was blown to the southern margin of the Channel. The main wind directions deduced from previous studies are N-NW in Brittany to NW in Normandy and Picardy (Lautridou, 1985), and close to N in Southern Belgium (Juvigné, 1985) (Figure 4).

The clay minerals mainly comprise smectite and vermiculite (often interstratified), except in Normandy, where kaolinite predominates, accompanied by a minor proportion of illite and chlorite (Lautridou, 1985). The biological content (pollen, moluscs, micromammals remains…) indicate a cold steppe environment.

Typical loess is characterised by (Smalley, 1971, 1975, Jamagne et al., 1981, Lautridou, 1985, Pécsi, 1990) :

- an homogeneous and porous structure
- the absence of bedding
- a sigmoidal cumulative grain-size distribution curve, with a straight central section indicating an abundance of coarse silt (20-50µm), and a median particle size between 25 and 35µm, a low sand fraction (1-2% max), 15-18% of clay (<2µm)
- the presence of primary carbonates (up to 15-18 %)
- up to 70 % of quartz grains, 1.5 to 2% of total iron, and less than 0.2% of organic carbon.

Pécsi (1990) proclaimed that "loess is not only a dust accumulation". Indeed, the loess microfabric observed by thin section, SEM and laser scanning, is characterised by weathering of primary calcite and dolomite grains followed by precipitation of calcareous cements as a result of a syngenetic "loessification" (Balescu, 1986; Derbyshire et al., 1998). The quartz silt grains are angular with occasionally chipped edges. Other important characteristics include : a random orientation of the grains, clay bridges between silt particles, rare clay coatings on silt grains, secondary carbonates on the grains and the biopores (Becze-Deack et al., 1997). Sand grains are subangular and show angular splinters with conchoidal faces, rare cupules, V-shaped impact cracks, triangular dissolution cavities, and often a coating of amorphous silica (Coudé and Balescu, 1987).

## 3 The collapse of loess

Collapse is a typical feature of unsaturated rather loose and low plastic soils, which are typical features of loess. Collapse is a significant volume reduction observed when wetting an unsaturated sample under load. This phenomenon has been described for a long time in arid regions. Jennings and Knight proposed in 1957 the double oedometer method to estimate the amplitude of possible collapse. Also, Gibbs and Bara proposed in 1962 a simple collapse criterion based on the dry density and liquid limit.

As commented by Houston (1995), there are numerous potential sources of water that can cause soil wetting :

- Broken water lines, canals and landscape irrigation (Walsh et al. 1993)
- Roof run-off and poor surface drainage (Walsh et al. 1993)
- Intentional and unintentional recharge (Shmuelyan 1995)
- Rising ground water table (El Nimr et al. 1995)

- Damming due to cut/fill construction (Kropp et al. 1994, Noorany and Stanley 1990)
- Moisture migration due to capillarity and protection from the sun (Jimenez-Salas 1995).

Li (1995) also mentioned collapse problems during impoundment of earthdams made up of poorly compacted loessial soils in China. Similar problems have been encountered by Lefebvre and Belfadhel (1989) in earthdam made up compacted non plastic till in the Baie James project in Canada.

Barden et al (1973) gave three conditions to observe collapse in a soil :

- an open potentially unstable partly saturated structure
- a high enough value of applied stress component to develop a metastable condition
- a high enough value of suction (or other bonding or cementing agent) to stabilise intergranular contacts, and whose reduction on wettong will lead to collapse

Simple capillary forces have often been mentioned as possible binding agent. However, as stated by Barden, the majority of collapsing soils involved the action of clay plates in the bonds between the bulky sand and silt grains. Possible effects of other chemical cementing agents like iron oxide or calcium carbonate are also mentioned.

Early scanning electron microscope (SEM) observations have been conducted by Barden et al. (1973) and Grabowska-Olszewska (1975) on European loess from Belgium and Poland. Grabowska-Olszewska (1975) provided a detailed description of the Golebice loess (8%<2μm) from the Baltic (Würm) Glaciation.

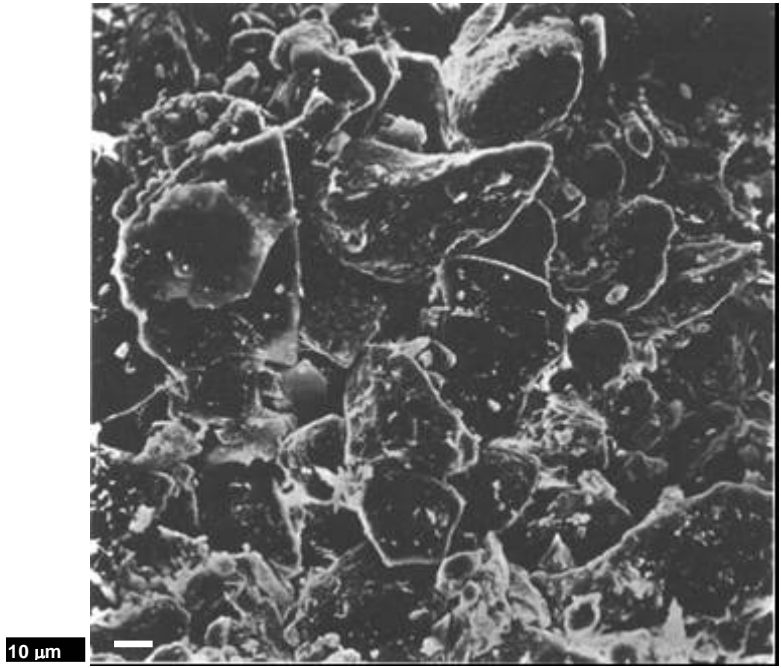

Figure 5. SEM observation of Golebice loess (Grabowska-Olszewska 1975).

As shown in Figure 5, Grabowska-Olszewska observed that the Golebice loess was characterised by a typical loose structure, some angular and sharp edged grains and some rounded ones (mostly quartz sand and silts grains), with fine particles often attached to the surface of larger grains. The fine particles were composed of clays minerals – mainly illite with some kaolinite – fine dispersional quartz, carbonates and iron oxides. The fine quartz particles observed were supposed to be syngenetic in relation with larger grains, also formed as a result of glacial grinding and submitted to aeolian transportation by being attached to larger quartz grains. The irregular depressions and cavities observed on the grain surfaces were related to weathering processes, whereas scratches and cracks were related to glacial grinding on parent rocks of loess. The uniform grain size distribution origin of the clay particles observed in this loess may indicate either aeolian transportation of in situ origin as a result of structural alteration of aluminosilicates occurring in the loess.

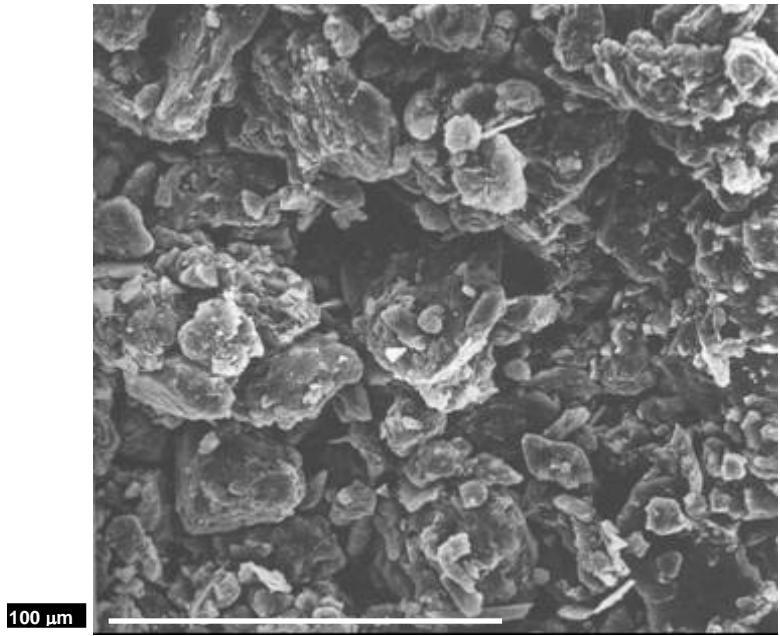

Figure 6. SEM observation (vertical section) of undisturbed Malan loess (Derbyshire et al. 1995).

The role of carbonate has been discussed by Grabowska-Olszewska according to the form of carbonate components observed using SEM. Since carbonate most often appeared under the shape of growths irregularly distributed in the structure, or less frequently as carbonate rock debris, the hypothesis of carbonate being a coating cementing agent linking the grains together in this Polish loess was contested. Conversely, based on SEM observations carried out on the Chinese Malan loess (Figure 6), Derbyshire et al. (1995) considered that, in the absence of any clay mineral, carbonate cementation was acting in this loess as a linking agent between particles. Note also in Figure 6 the loose structure and large voids (up to 10 μm in diameter), the size of the grains (from 5 to 20 μm) and their angular shape.

Figure 7 (Derbyshire et al. 1995) shows an impressive large sinkhole observed in Malan loess near Lanzhou.

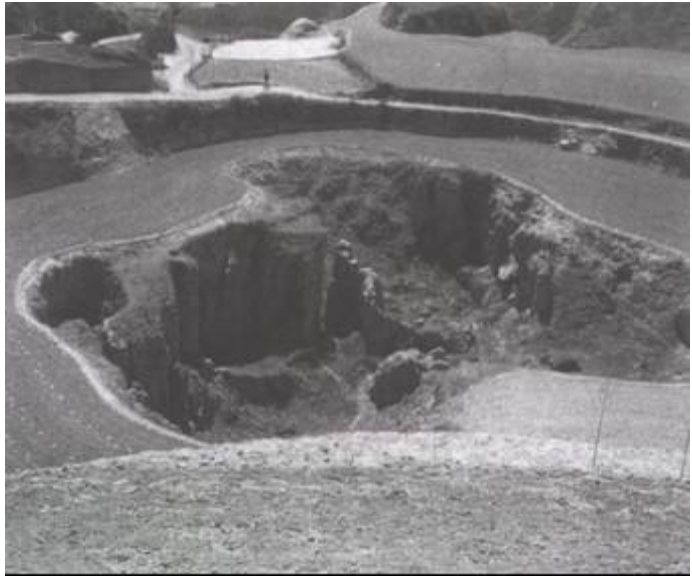

Figure 7. Large (30 m diameter) sinkhole in Malan loess near Lanzhou (Derbyshire et al. 1995)

### 3.1 Application of the mechanics of unsaturated soils to collapse

As commented above, Jennings and Knight (1957) proposed a simple test, called double oedoemeter test, to characterise the collapse susceptibility of loess. The principle of the test is described in Figure 8. It consists in running two oedometer compression tests :

- a test at a constant initial water content, corresponding to the initial natural state of the soil
- a test in which the unsaturated sample is wetted under a small load, and subsequently loaded in a zero suction condition, close to saturation.

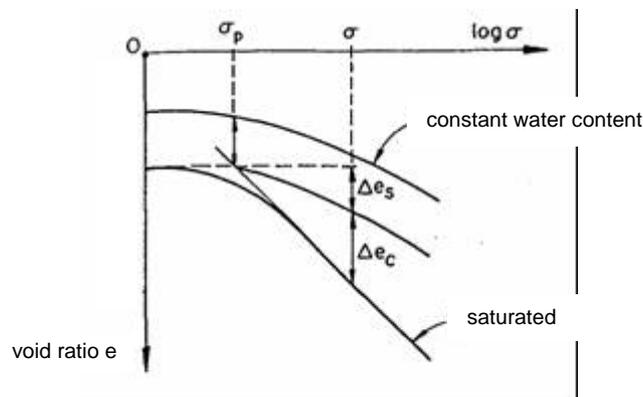

Figure 8. The double oedometer method (Jennings and Knight 1957).

The amount of collapse as a function of the applied stress is taken from the difference of void ratio taken from the two curves under the corresponding stress. Note that the two experimental curves may have to be initially fitted at low stresses to account for the natural variability of natural loess.

Indeed, the collapse phenomenon was the main objection initially made by Jennings and Burland (1962) to the proposal of an effective stress extended to unsaturated soils, as made by Bishop (1958). It was shown that the decrease of suction occurring during water flooding under constant stress would induce a decrease of Bishop'stress ($\sigma' = \sigma - u_a + \chi (u_a - u_w)$, with $0 < \chi < 1$) that should result in a swelling strain of the soil.

The double oedometer method, that only considered either the unsaturated state at a constant initial water content or the zero suction and saturated state of the soil after waterflooding has later been extended and confirmed by experimental tests carried out under controlled suctions on samples of compacted silt by Matyas and Radhakrisna (1968). The principle of the double oedometer method is in fact compatible with the notion of state surface proposed by Matyas and Radhakrisna and presented in Figure 9.

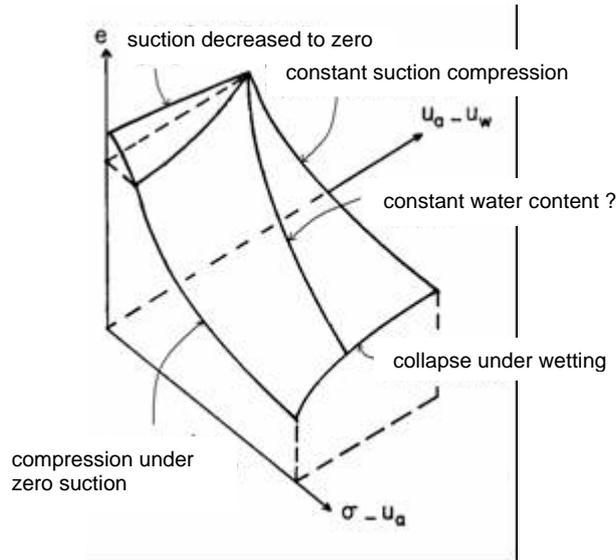

Figure 9. The state surface concept (after Matyas and Radhakrisna 1968).

In this Figure, Matyas and Radhakrisna proved the existence of a state surface that characterised the volume change behaviour of unsaturated soils submitted to increased degree of saturation (either by loading or by decreasing suction) in a $e / (p - u_a) / (u_a - u_w)$ 3D space [being respectively $e$ the void ratio, $(p - u_a)$ the net mean or vertical stress and $(u_a - u_w)$ suction]. The existence of the state surface proves :

- that the more compressible state of the sample is the flooded state under a zero suction (that actually does not necessarily coincides with full saturation of the sample, due to entrapped air bubbles)
- that the suction decrease curves will converge towards the zero suction compression curve. A particular path leading to this curve is followed when carrying out the double oedometer method.

It has for a long time been considered that suction was decreasing when loading an unsaturated soil at a constant suction. Actually, experiments carried out on compacted soils showed that this was only true when coming close to saturation, as seen in Figure 10 (Li 1995, Gens et al. 1995). In this figure where the saturation hyperbola is also represented, constant water content compression tests correspond to vertical upwards straight lines. Suction measurements carried out at various positions in the $(\gamma_d / w)$ plane (being $\gamma_d$ and $w$ respectively the dry unit weight and the water content) show that no suction change was observed as far as the degree of saturation of the soil is not too high and smaller than around 85%. The interpretation of this behaviour was based on the fact that dry compacted soils are composed of aggregates with air filling the inter-aggregates pores (see Delage et al. 1996). Hence, suction is governed by intra aggregates phenomena. No suction changes should occur when macroscopic compression only reorganises the microfabric by getting a denser assemblage of aggregates without compressing the aggregates themselves. This is apparently the case in the dry states, where the aggregates are rigid enough to break or change in shape during compression, without any compression of the aggregates themselves.

The question is now how far these results apply to collapsible loess. Looking at the SEM photos shown in Figure 6, it appears that large inter-grains pores are most probably full of air and that suction effects predominantly concern smaller pores. As in dry compacted soils, it is suspected that compression will collapse the large inter-grains pores and probably not affect the smaller pores. Like in dry compacted soils and due to their low initial degree of saturation, it seems likely that constant water compression test in collapsible loess would occur at a constant suction. Hence, the constant water content compression test followed during Jennings and Knight's test would be part of the state surface, together with the zero suction compression line.

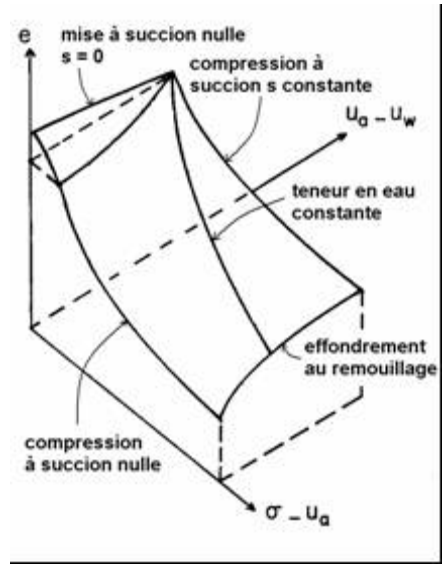

Figure 10. Suction isovalues of a compacted soil (Li 1995)

Being irreversible, the collapse strain is by definition a plastic strain. In this regard, the concept of Loading-Collapse (LC) yield curve proposed by the Barcelona group (Alonso et al. 1990) was a fundamental contribution since it provided for first time the integration of the collapse strain within an elasto-plastic framework, as shown in Figure 11. The curved shape of the LC curve in the ($s$ / $p - u_a$) plane (being $s$ and $p - u_a$ respectively the suction and the mean net stress) allows for the mobilisation of the LC yield curve when reducing suction from a value $s_3$ down to zero under a constant load (vertical downwards stress path). The yield curve displacement results in a hardening plastic mechanism that corresponds to a plastic compression of the sample. Based on a given set of constitutive parameters mostly taken from the volume change behaviour under various constant suctions, the Barcelona basic model (BBM) hence allows for the calculation of collapse as a plastic strain.

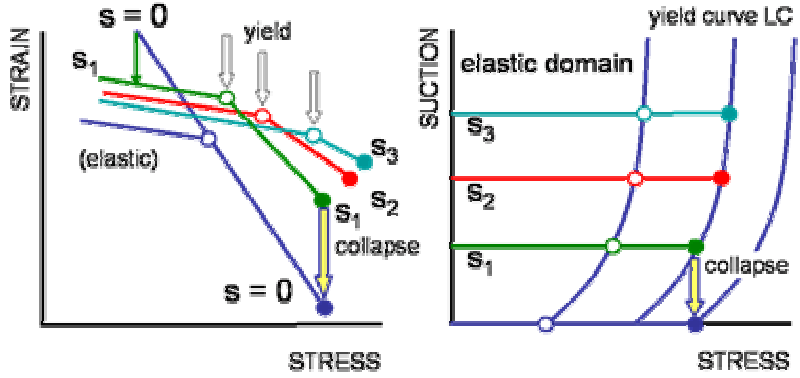

Figure 11. Collapse as a plastic mechanism : the LC curve (Alonso et al. 1990)

Besides the elasto-plastic volume change framework described above, the BBM model also considered the constitutive behaviour in the $q / p – u_a$ using the Cam Clay model to account for the effects of deviator stresses. To simplify, Alonso et al. (1990) proposed to use the modified Cam Clay ellipse as a yield curve, the size of the ellipse in the $q / p – u_a$ plane being enlarged by moving the extended "preconsolidation" pressure by following the LC curve in the $s / p$ plane when suction is increased.

Actually, based on a possible similarity between constant water content and constant suction curves, it appears interesting to consider the results of triaxial tests performed by Lin (1995) and shown in Figure 12. Tests have been performed at various stress inclinations K ($K = \sigma_3 / \sigma_1 = 0.3$, 0.5 and 0.7). Yield values have been taken at a value of 1.5% of the volumetric strain, at which curve all presented a significant change in slope considered as being representative of yield. The standard constant confining pressure triaxial tests exhibited a fragile behaviour with noticeable peaks observed in the ($q / \varepsilon_1$) stress-strain curves.

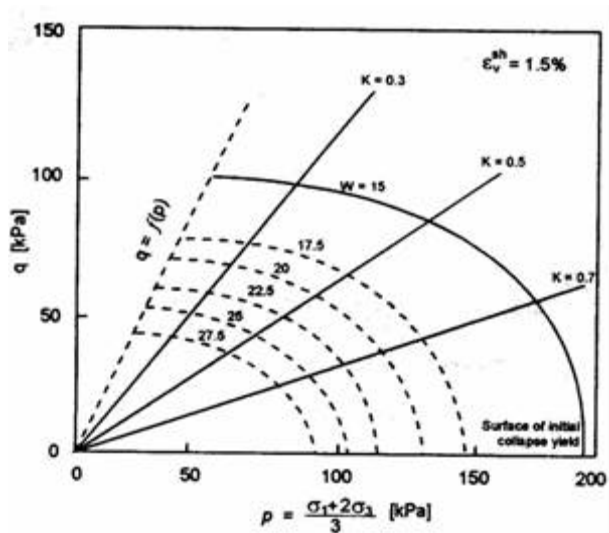

Figure 12. Yield curves of an undisturbed Chinese loess (Lin 1995)

The shape of the yield curves are interestingly similar to the ellipse considered in the Cam Clay model adopted in the BBM model, except that it is cut by frictional rupture, as in various cap models. Yield stresses under isotropic compressions tests increase with decreased water contents, i.e. increased suction. This suction hardening phenomenon is modelled by the LC curve. The water retention curve of the loess tested is however not provided in the paper. It would have allowed an estimation of the suctions corresponding, in the initial state, to the water contents used here. The values of suctions would have allowed the determination of the LC curve. The symetric shape of the ellipse versus the *p* axis shows that the mechanical response of the loess is isotropic.

The set of results extracted here from various publications show that the present state of the art of the mechanics of unsaturated soils helps better understand and model the behaviour of loess. Note however that the peak effect and the sudden and significant collapse observed in some cases should ask for a better refinement in the constitutive modelling, by introducing also the effects of cementation and the impact of structure effects on the behaviour.

## 4. Study of the loess along the TGV railways

### 4.1 Typical loess profiles along the TGV line
The section of the railways in which water sensitivity off loess deposits crossed is presented in Figure 4. The three profiles are denoted as CH (Chaulnes), LT (Le Transloy) and CR (Croisilles). A detailed description of the various facies that have been observed in these profiles has been made, as shown in Figure 13.

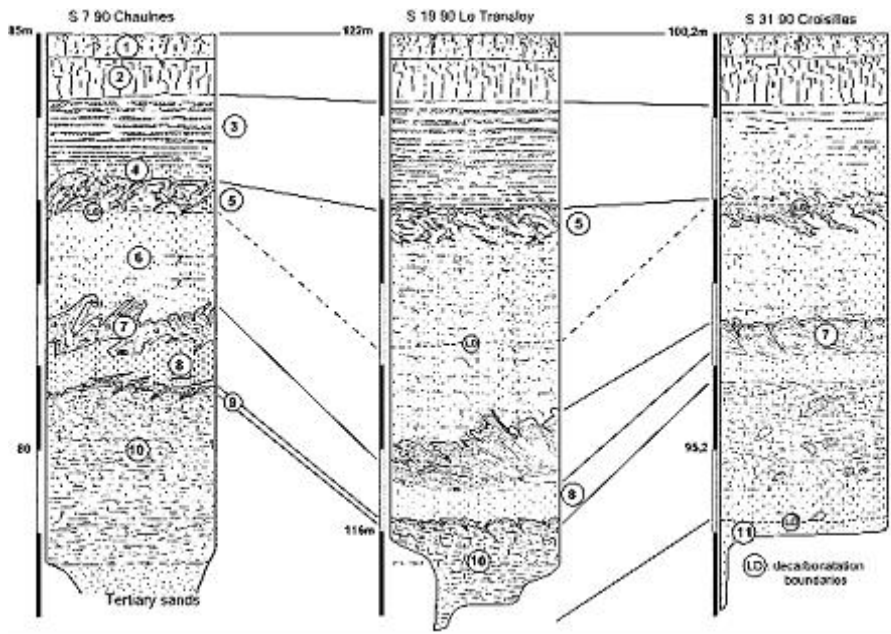

Figure 13. Examples of loess profiles observed along the Paris Lille TGV line

1 – Ploughing horizon (Ap) of he surface soil, 2 – Clayey horizon (Bt) of the surface brown leached soil (Luvisol), 3 – Doublet horizon of the surface brown leached soil (banded), 4 - Decalcified loess, 6 - Finely laminated calcareous loess with small cryodessication frost cracks, 5, 7 and 9 – Tundra gley horizons with cryoturbations (5&7), and (or) solifluction (9) features. At Le Tranloy, a bone remain from a large mammal (Bison) was found 8 – Homogeneous calcareous loess with very fine (<2mm) foliated structure (Freeze-Thaw pattern), 10 - Bw horizon of Boreal / Arctic brown soils (Cambisols), with medium (4-6mm) foliated structure (Freeze-Thaw pattern), 11 – Calcareous loess

### 4.2 Geotechnical characterisation of the samples

Soil samples have been manually extracted in September 2002 at various depths in a site located at 140 km North of Paris, at a 25 m distance of the TGV line, not far from the site of Croisilles (CR in Figure 4). Four block samples have been trimmed and cut by using a shovel at different depths (1.2, 2.2, 3.5 and 4.9 m) in a 1.5 m large and 9 m long trench that was excavated step by step. Visual observation showed that samples were brown (1.2 and 4.9 m) or light brown (2.2 and 3.5 m). X ray diffractometry showed that the samples at 1.2, 2.2 and 3.5 m were mainly composed of quartz and feldspar with some clay. Some carbonates (calcite and dolomite) were observed in the 4.9 m sample. Analysis carried out on the clay fractions (< 2 µm) showed that in all cases the clay fraction comprised kaolinite, illite and interstratified illite-smectite. Figure 14 shows that the grain size distributions curves of four soils are comparable and typical of loess, with a slightly higher clay fraction (18 and 20%) in the 1.2 and 4.9 m samples respectively, as compared to 16% for the two other samples.

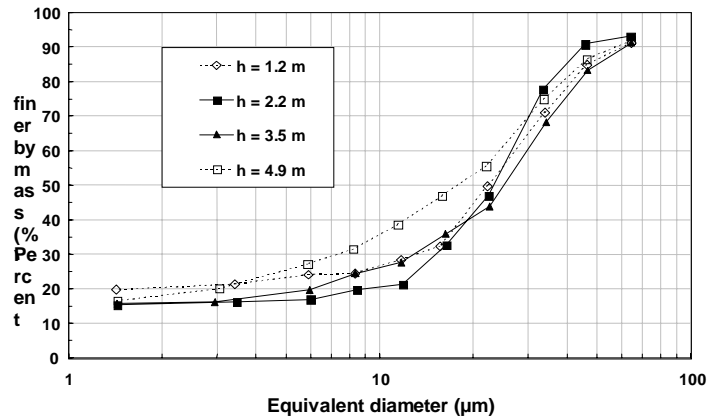

Figure 14. Grain size distribution curves of the four samples

Table 1. Geotechnical properties of the four samples

| Depth (m) | Horizon | $\rho_s$ (Mg/m³) | % < 2µm | $w_L$ (%) | $w_p$ (%) | $I_p$ | $\rho_d$ (Mg/m³) | $w_{Nat}$ (%) | $S_{rNat}$ (%) | %Ca | Suction (kPa) |
|---|---|---|---|---|---|---|---|---|---|---|---|
| 1.2 | 3 | 2.719 | 20 | 30 | 21 | 9 | 1.52 | 18.9 | 66 | 10 | 20 |
| 2.2 | 6 | 2.714 | 16 | 28 | 22 | 6 | 1.39 | 18.1 | 53 | 12 | 34 |
| 3.5 | 7 | 2.713 | 16 | 26 | 20 | 6 | 1.54 | 16.6 | 55 | 11 | 27 |
| 4.9 | 10 | 2.712 | 18 | 30 | 21 | 9 | 1.55 | 23.7 | 82 | 16 | 14 |

Table 1 presents the geotechnical characteristics of the profile, including the degree of saturation $S_r$ and the in-situ suction measured using the filter paper method (Fawcett & Collis-George 1967). For each depth, the number of the corresponding horizon described in Figure 13 is indicated.

It is observed that :

- the solid unit mass $\rho_s$ is almost the same at all depths
- plasticity indexes $I_p$ are quite small, ranging from 6 to 9
- the dry unit mass is also low with $\rho_d$ varying from 1.39 to 1.55 Mg/m³
- the degrees of saturation $S_r$ are low (less than 66%) in samples at 1.2, 2.2 and 3.5 m
- the carbonate content is higher than 10%, reaching 16% for the sample at 4.9 m
- suction is small in all samples, comprised between 14 and 34 kPa.

The sample at 2.2 m depth has the smaller clay fraction, the smallest plasticity index, the lowest density and degree of saturation and the highest suction. This sample is probably the more collapsible.

Scanning Electron Microscope photos taken on freeze dried (Tovey & Wong 1973) and freeze fractured samples taken at 2,2 m are shown in Figure 15 to 17. Samples were ultra rapidly frozen by plunging small pieces of loess (2 mm wide and 5 mm long sticks) in nitrogen previously cooled down to its freezing temperature by applying vacuum. No nitrogen boiling is observed in such conditions, allowing for a very fast freezing with no ice expansion and satisfactory microstructure preservation. Freeze fracturing is an interesting technique beacuse the particles are strongly held together when fracturing the sample. Hence, the fracture is determined by ice failure and it crosses through the different levels of structures without being influenced by the structure itself.

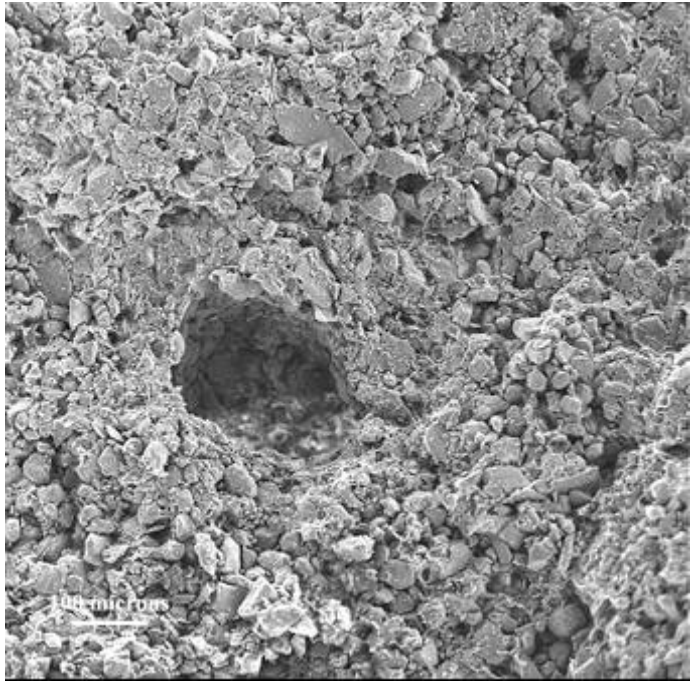

Figure 15. SEM observation of the 2.2 m soil, overall view.

The overall view of the structure presented in Figure 15 shows a significant but irregular presence of agglomerated clay particles within a skeletal assemblage of silt grains (around 20-30 µm in diameter). In areas were they are present, aggregated clay platelets fill the inter-grains pores. In other areas, the silt grains appear clean with no clay coatings around them. Obviously, clay acts as a link in the areas where it is present. The clean grain zones with larger pores (around 20 µm in diameter) probably correspond to weaker zones that present a local volume decrease larger than in clay rich zones. Collapse is then probably localised in such zones where pores as large as 10 µm can be observed. Similar local heterogeneity under macroscopic volume changes has also been observed in sensitive clays where the compressibility is such that the term collapse is also used to describe the sudden decrease in volume observed once the yield stress is passed during one dimension compression (Delage & Lefebvre 1984). Note also the large pore (200 µm in diameter). This pore can correspond to the cavity remaining after a sand grain has been extracted during freeze fracturing, as suggested by the smooth surface aspect observed inside the cavity.

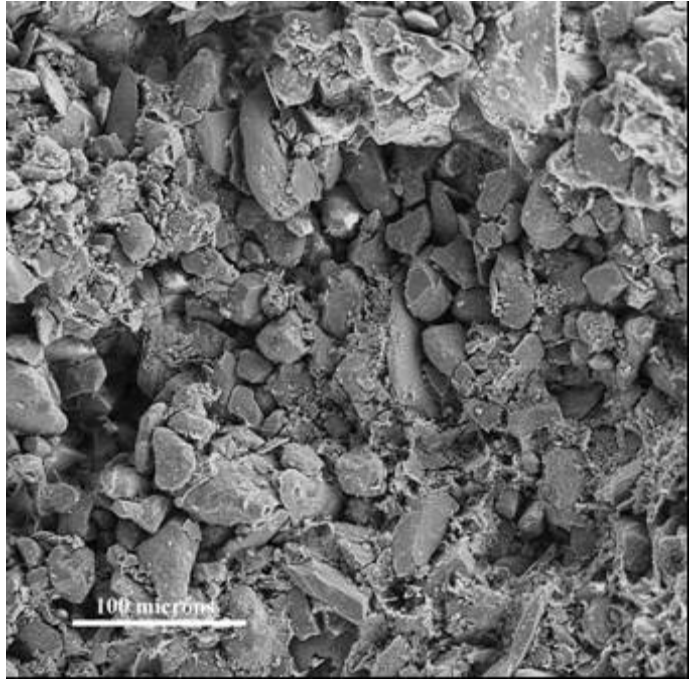

Figure 16. SEM observation of the 2.2 m soil, closer view.

Figure 16 shows a closer view giving more details on the shape of the irregular, angular and sharp edged grains that are most probably composed of quartz. The filling of some inter-grains pores by agglomerated clay particles is also observed more in details thus clearly illustrating the significant link provided by clay between the silt grains in these zones. Between the grains, the significant difference between clay filled pores and voids is also more apparent. Few clay coatings appear on the grains in the areas with no clay.

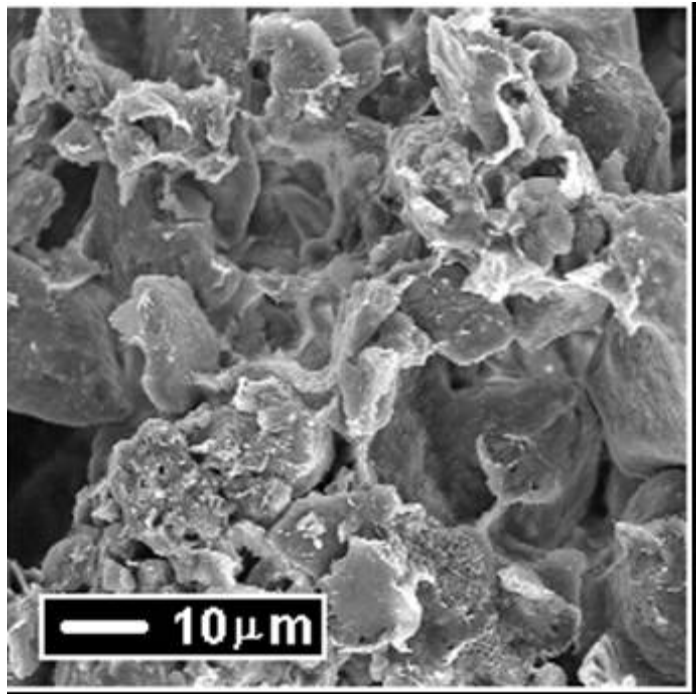

Figure 17. SEM observation of the 2.2 m soil, detail of clay-silt interaction.

Figure 17 presents, in a clay rich zone, the details of the contact between clay aggregations and silt grains, with apparent clay bridges and obvious inter-grains linking action.

## 5. Collapse susceptibility of the Croisilles samples

### 5.1 Collapse susceptibility of samples at natural water
The collapse susceptibility of the samples was firstly characterised at natural water content using Knight (1963)'s simple oedometer method, as seen in Figure 18.

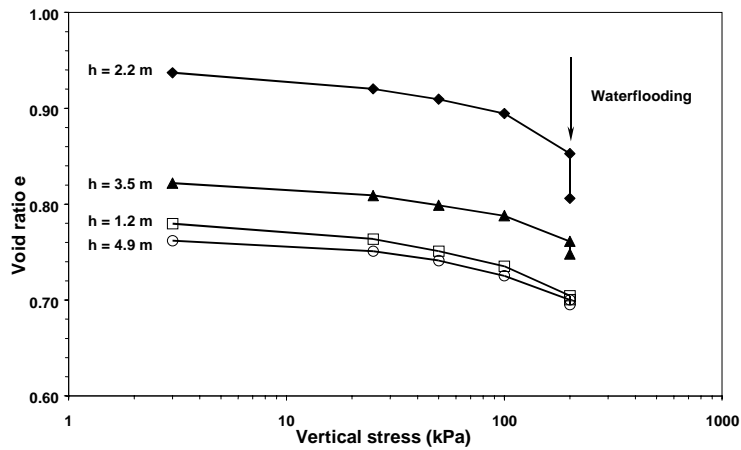

Figure 18. Collapse test, simple oedometer method

The method consists in loading the sample at natural water content up to 200 kPa and then soaking the sample under 200 kPa. According to the volume decrease observed during this test, various degrees of susceptibility starting from non collapsible (0 – 1%) to very collapsible (10 – 20%) have been defined by Knight, as presented in Table 2.

Table 2. Knight's Criterion (1963)

| Collapse strain under 200 kPa (%) | Susceptibility | Evaluation |
| --- | --- | --- |
| 0 – 1 | non collapsible | Samples at 1.2, 3.5, 4.9 m |
| 1 – 5 | slightly collapsible | Sample at 2.2 m |
| 5 – 10 | Collapsible | |
| 10 – 20 | very collapsible | |

At each depth, a 20 mm high and 70 mm diameter specimen was prepared by using a ring sampler. The results in Figure 18 clearly show that only the sample at 2.2 m exhibits some collapse (2.41%), whereas samples at 1.2 m, 3.5 m and 4.9 m show negligible collapses (0.23% for 1.2 m, 0.72% for 3.5 m and 0.28% for 4.9 m). In Knight's classification (see Table 2) 2.41% corresponds to a slightly collapsible soil (between 1 and 5%).

Microstructure changes due to wetting were studied on the four samples by SEM observation and mercury intrusion pore size distribution measurements on freeze dried samples (Tovey and Wong 1973). A significant change in the pore size distribution (PSD) curve was only observed in the 2.2 m sample, as shown in Figure 19.

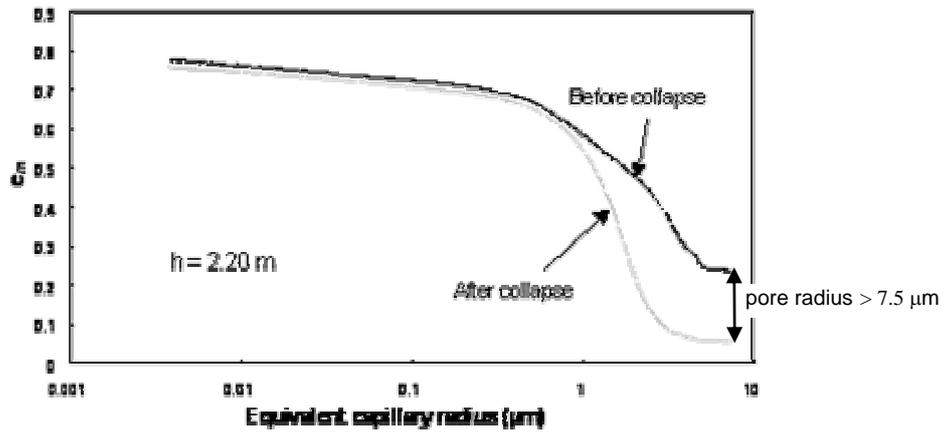

Figure 19. Pore size distribution curves of the soil 2.2 m, before and after collapse under 200 kPa stress

In this Figure, the pore volume (y axis) is expressed in terms of void ratio whereas the x axis represents the apparent entrance radius of pores, derived from the applied mercury pressure, under the hypothesis of cylindrical pores (Diamond (1970). The curves correspond to pore size that can be investigated by the porosimeter i.e. pore sizes included between a maximum entrance radius of 7.5 µm (atmospheric pressure, on the right hand side) and a minimum entrance radius of 3.5 nm (highest applied mercury pressure equal to 20 MPa, left hand side). The pore volume corresponding to radii larger than 7.5 µm was determined by measuring the volume intruded when passing from vacuum to atmospheric pressure. The precise measurement of the total volume of the sample gives the total pore volume. It allows the determination of the remaining porosity with entrance pore radii smaller than 3.5 nm. Based on that, it can be observed in Figure 19 that the void ratio corresponding to entrance pore radii larger than 7.5 µm before collapse is equal to 0.24.

It clearly appears that the collapse due to wetting corresponds to i) a decrease of large pores greater than 7.5 µm, and ii) a decrease of medium sized pores between 0.7 and 7.5 µm, with no effect on smallest pores, below 0.7 µm. Also, the dominant large pore population after collapse is well graded around an average value of r = 1.8 µm. Referring to the SEM photos previously presented, the collapse apparently affects the larger inter-grains pores and results in a more organised and homogeneous microstructure characterised by a well graded pore population around 1.8 µm. The smaller porosity not affected by soaking corresponds to the porosity inside the clay aggregations that were observed as beinf irregularly scattered within the skeletal granular structure made up of the silt grains.

SEM observations on collapsed samples however did not show any significant microstructure changes, even in the sample at 2.2 m. However, some differences were suspected in the appearance of the clay fraction, with a smoother appearance of the hydrated clay platelets, as compared to the drier clay platelets observed before soaking (Delage et al. 1996).

### 5.2 Effect of water content on the collapsibility

Since the natural water content in the loess profile was suspected to change due to seasonal effects, the effect of initial water content on collapse was investigated on the 2.2 m sample. The collapse susceptibility was investigated at 6 different water contents : 0, 4, 10, 14, 18% (natural water content), and at 23%. The starting water contents were obtained either by drying soil in the laboratory atmosphere for a given time ($w < 18\%$) or by adding water using a wet filter paper ($w > 18\%$).

The simple and double oedometer methods (Jennings & Knight 1957) were both used. In the double oedometer method, two oedometer tests were conducted in parallel: a compression test under constant water content up to 800 kPa and a compression test up to 800 kPa on a sample previously saturated under 3 kPa. Figure 20 shows the results obtained with $w_i = 4$, 14 and 23%. In all tests, a satisfactory agreement is observed between the two methods, confirming the validity of the double oedometer method. No fitting between the different curves was necessary.

Clearly, collapse increases with decreased initial water content, as synthetically shown in Figure 21 which also reports the data obtained at the three other water contents ($w = 0$, 10 and 18%). The good agreement between the two methods is observed at all water contents and the decrease in collapse with increased water content appears to be significant and roughly linear. Significant collapse susceptibilities are observed in drier states, showing the need of properly monitoring the changes in natural water content profiles resulting from ground-atmosphere exchanges (Blight 1997), particularly during the drier ant wetter periods.

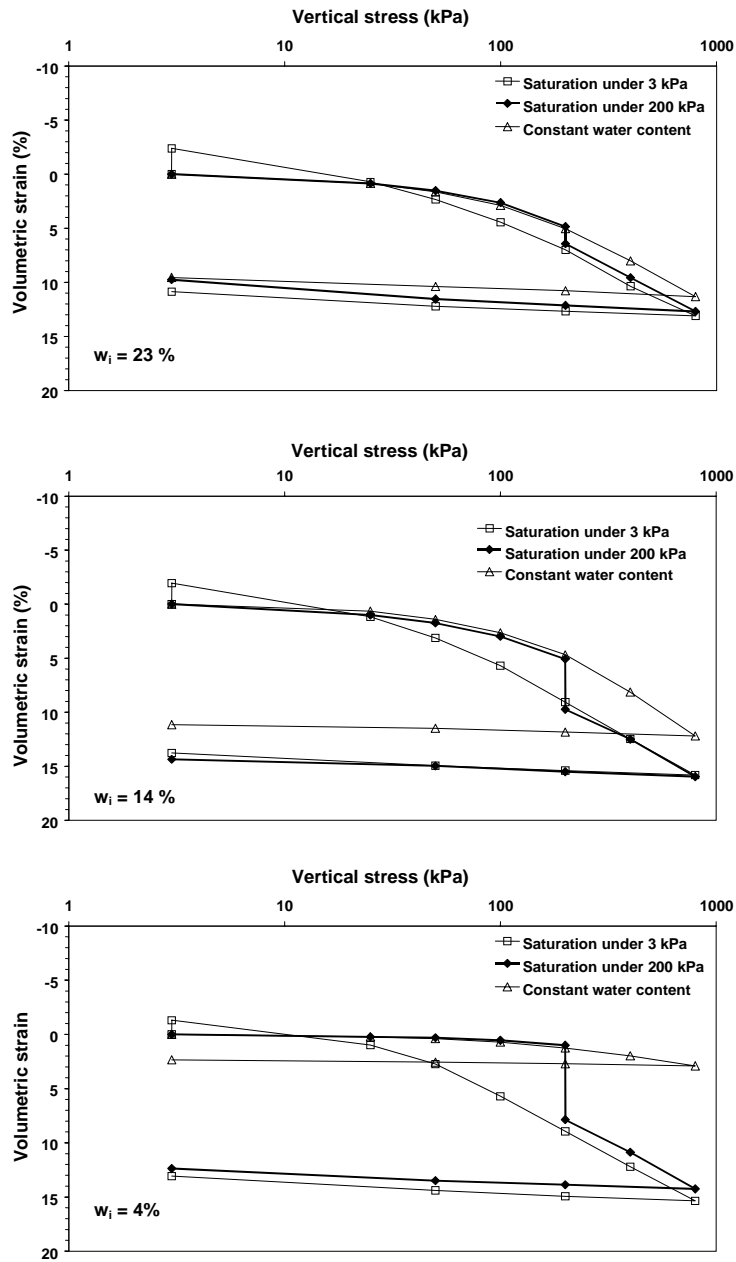

Figure 20. Determination the collapsibility of soil 2.2 m by using simple and double oedometer methods, $w$ = 23, 14 and 4 %

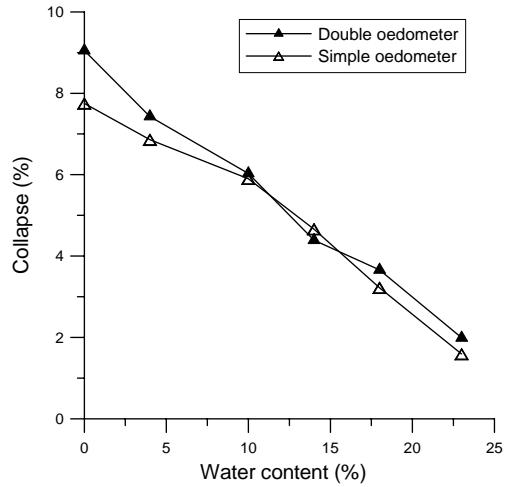

Figure 21. Collapse of soil 2.2 m at different initial water contents, identified by simple and double oedometer methods.

## 5.3 Evaluation of various collapsibility criteria

As seen previously, the best way to assess collapse susceptibility of soils is based on the use of the oedometer, using either the simple or the double oedometer method. However, a simpler determination based on standard geotechnical parameters would obviously be satisfactory and interesting, particularly for long linear infrastructures. Number of authors have proposed such criteria, that are now examined on the samples at natural water content, and compared to oedometer results.

As seen before, estimation according to Knight's criterion (1963) lead, for the 2.2 m deep sample, to the grading "slightly collapsible".

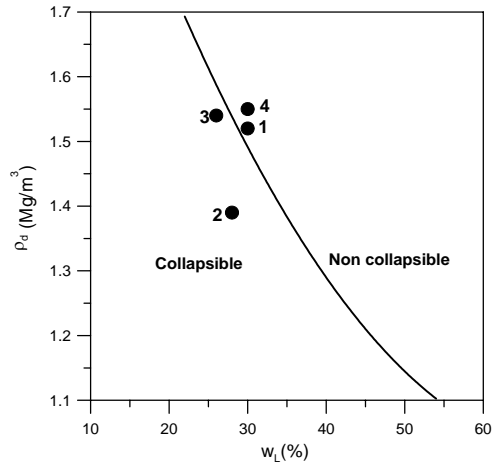

Figure 22. Collapsibility of the four soils according to Gibbs and Bara (1962)

Gibbs and Bara (1962) proposed a criterion based on the values of the dry unit mass $\rho_d$ and of the water content $w$ of the soil. They considered that any soil having a dry unit mass high enough to achieve upon saturation a water content equal or higher than the liquid limit would be collapsible. Figure 22 shows the graphic representation of this criterion together with the points corresponding to the samples tested here. It shows that the 1.2 m and 4.9 m samples are considered as non collapsible whereas the 2.2 m and 3.5 m samples are considered as collapsible. Since the 2.2 m sample is located further from the separation line than the 3.5 m sample, it is more collapsible.

Other less known criteria, presented and used in Huergo et al. (1989) are now considered. As Gibbs & Bara's criterion, these criteria are mainly based on characteristics of density and plasticity. The relevant geotechnical parameters for all criteria are those presented in Table 1, completed by those of Table 3 regarding the values of void ratio. Table 3 also gives the values of the relevant parameters of each criterion.

Table 3. Results of various collapse criteria

| Depth (m) | $e_0$ | $e(w_L)$ | $K_e$ Denisov | $w_p-w_{Nat}>0$ Priklonskij | $K_L$ Feda | $\delta$ (%) Stephanoff & Kremakov | $\delta$ (%) measured |
|---|---|---|---|---|---|---|---|
| 1.2 | 0.780 | 0.815 | 1.04 | 2.1 | 0.85 | 3.37 | 00.23 |
| 2.2 | 0.937 | 0.760 | 0.81 | 3.9 | 2.03 | 5.00 | 2.41 |
| 3.5 | 0.822 | 0.705 | 0.86 | 3.4 | 1.70 | 3.48 | 0.72 |
| 4.9 | 0.762 | 0.814 | 1.07 | -2.7 | 0.88 | 1.66 | 0.28 |

Denisov' criterion (1951, in Huergo et al. 1989) is based on a parameter $K_e$ obtained when comparing the void ratio corresponding to the liquid limit $e(w_L)$ to the natural void ratio $e_{Nat}$, as follows : $K_e = e(w_L)/e_{Nat}$. A soil is collapsible if $K_e < 0.75$. Observation of Table 3 shows that none of the samples is collapsible according to this criterion.

According to Priklonskij (1952, in Huergo et al. 1989), a soil is collapsible if its natural water content is larger than its plastic limit, giving $w_P - w_{Nat} < 0$. Data in Table 3 show that it is only the case of the 4.9 m sample.

Feda (1964, in Huergo et al. 1989) proposes a criterion based on a parameter function of the natural water content, of the degree of saturation, plastic limit and plasticity index $K_L = [(w_{Nat}/S_{rNat}) - w_P] / I_P$. According to Feda, a loess is collapsible if $K_L > 0.85$ and $S_{rNat} > 0.6$. Table 3 shows that all samples are collapsible according to this criterion.

The approach of Stephanoff and Kremakova (1960) is more ambitious since it proposes a criterion allowing for a quantitative estimation of the collapse. The criterion is based on the natural porosity n and the natural water content, together with a coefficient K which accounts for the soil type ($K = 0.08$ for a clayey loess and 0.05 for a loess). The collapse strain $\delta$ is obtained by the following relation : $\delta = K (n-40) (30-w_{Nat})$. The calculations of $\delta$ in Table 3 were made with $K = 0.08$ for samples at 1.2 and 4.9 m and $K = 0.05$ for samples at 2.2 and 3.5 m. The comparison between calculated and measured values shows that this criterion significantly overestimates collapse strain.

Examination of the previous criteria and comparison with the oedometer tests show that only the Gibbs & Bara criterion gives a reasonable estimation of the collapse susceptibility of the loess tested here. In particular, it seems not possible to obtain a quantitative estimation of collapse based on standard geotechnical parameters, as proposed by Stephanoff and Kremakova (1960).

## 6 Conclusions

A geological description of the nature, of the repartition and of the origin of loess deposits in Northern France has been presented to better characterise some problems related to the collapse susceptibility observed in some areas that are crossed through by the TGV Nord railway. Special attention was also focused on microstructure features of loess sediments. The collapse behaviour of unsaturated soils has then been replaced within the framework of unsaturated soil mechanics, showing that collapse could be considered as a plastic deformation in the elasto-plastic framework provided by the Barcelona group (BBM model).

An investigation of the collapse susceptibility of four intact samples extracted from a profile located along the TGV line, 140 km far from Paris has been conducted by using the simple and double oedometer methods. The most sensitive sample of the profile presented a slight collapse susceptibility at the natural initial water content that corresponded to the sampling conditions. Collapsibility appeared to be significantly sensitive to changes in water content with collapse volume decrease higher than 5% at water contents smaller than 12%. This increased collapsibility may explain the sinkholes observed in areas where collapsible soil was exposed to atmosphere with possible drying occurring before soaking by rainfalls. This confirms the necessity of protecting the layer of sensitive loess during geotechnical works.

SEM observations showed a porous microstructure with heterogeneous scattering of clay aggregations that filled the inter-grains pores and worked as a linking agent between the grains in some areas. In areas with no clay, sharp edge angular silt grains (15 to 30 µm in diameter) were observed with large inter-grains pores. These pores located in the areas with no clay probably significantly contribute to collapse volume decrease. Mercury intrusion porosimetry identified the changes in inter-grains pores that occurred during collapse and showed that the smaller pores inside the clay aggregations were not affected. The collapse structure is apparently more organised with a well graded pore size distribution curve.

A examination of the relevance of various existing collapse criteria (Denisov 1951, Priklonskij 1952, Stephanoff and Kremakov 1960, Gibbs & Bara 1962 and Feda 1964) was carried out. Although all criteria detected the higher susceptibility of the 2.2 m sample, the comparison showed that the well known Gibbs & Bara's criterion provided the best answer, in a good agreement with the oedometer tests.

## Acknowledgements

The authors are grateful to Réseau Ferré de France and to the French railways company SNCF for its financial support and interesting scientific exchanges.